# Adoption of Artificial Intelligence in Schools: Unveiling Factors Influencing Teachers' Engagement

Mutlu Cukurova[1], Xin Miao[2] and Richard Brooker[2]

[1] University College London, London, UK.
[2] ALEF Education, UAE.
m.cukurova@ucl.ac.uk

**Abstract.** Albeit existing evidence about the impact of AI-based adaptive learning platforms, their scaled adoption in schools is slow at best. In addition, AI tools adopted in schools may not always be the considered and studied products of the research community. Therefore, there have been increasing concerns about identifying factors influencing adoption, and studying the extent to which these factors can be used to predict teachers' engagement with adaptive learning platforms. To address this, we developed a reliable instrument to measure more holistic factors influencing teachers' adoption of adaptive learning platforms in schools. In addition, we present the results of its implementation with school teachers (n=792) sampled from a large country-level population and use this data to predict teachers' real-world engagement with the adaptive learning platform in schools. Our results show that although teachers' knowledge, confidence and product quality are all important factors, they are not necessarily the only, may not even be the most important factors influencing the teachers' engagement with AI platforms in schools. Not generating any additional workload, increasing teacher ownership and trust, generating support mechanisms for help, and assuring that ethical issues are minimised, are also essential for the adoption of AI in schools and may predict teachers' engagement with the platform better. We conclude the paper with a discussion on the value of factors identified to increase the real-world adoption and effectiveness of adaptive learning platforms by increasing the dimensions of variability in prediction models and decreasing the implementation variability in practice.

## 1 Introduction

AI in Education literature has an increasing amount of research that shows the positive impact of using AI in adaptive learning platforms to support students' academic performance (VanLehn, Banerjee, Milner, & Wetzel, 2020), their affective engagement (D'Mello et al., 2007), and metacognitive development (Azevedo, Cromley, & Seibert, 2004) in controlled studies. Several learning platforms including OLI learning course (Lovett, Meyer, & Thille, 2008), SQL-Tutor (Mitrovic & Ohlsson, 1999), ALEKS (Craig et al., 2013), Cognitive Tutor (Pane, Griffin, McCaffrey, & Karam, 2014) and ASSISTments (Koedinger, McLaughlin, & Heffernan, 2010), have also been shown to have statistically significant positive impacts on student learning in real-world settings. It is important to note that some of these are conducted as multi-state studies using matched pairs of schools across the USA and still showed positive impact (i.e Pane *et* *al.*, 2014). These results are particularly significant, as more general studies examining the positive impact of educational interventions are notoriously hard to reach statistical significance (see also, Du Boulay, 2016). Despite such strong impact results, the adoption of adaptive AIED platforms in schools is slow in many schools at best, or simply doesn't exist in many others across the globe. There is increasing concern about how to promote the adoption of AI-based adaptive learning platforms in schools to enhance students' learning performance.

The slow adoption of AIED systems in real-world settings might, in part, be attributable to the frequent neglect of a range of other factors associated with complex educational systems. These include but are not limited to understanding and deliberately considering the learners' and the teachers' preferences (e.g. Zhou et al., 2021), why and how exactly in the system the tool will be used by the teachers (e.g. Buckingham Shum et al., 2019), the social contexts in which the tools will be used and the perceived/actual support offered to teachers from such social contexts in their practice, physical infrastructure, governance, and leadership of the schools (Banavides *et al.*, 2020), as well as the ethical (e.g. Holmes *et al.*, 2021) and societal implications related to fairness, accountability and transparency of the system (e.g. Sjoden, 2020). Within this complex ecosystem of factors, existing research tends to ignore the relevance of such holistic factors or focus mainly on investigating individual teacher-level factors with generic technology adoption considerations (e.g perceived usefulness and ease of use (Davis, 1989)). This teacher-centric way of looking at the educational ecosystems is useful but partial in that it ignores where the power lies in introducing such tools into educational institutions in the first place, be it government or local authority policy, the leaders of the school, senior teaching staff in the institution, school infrastructure and culture, community and collaboration opportunities among individual teachers or even the learners themselves. Besides, generic technology acceptance evaluations like Technology Acceptance Model tend to overlook the peculiarities of a specific technology like AI which has different implications on users' agency, accountability and transparency of actions. Likewise, they miss the issue of who builds the tools and thus the role of the technical, social and market forces within which educational technology developers must operate. Perhaps, this is, at least to a certain extent, due to the limited scope of AIED solutions focusing mainly on the technical and pedagogical aspects of delivery in a closed system, rather than taking a "mixed-initiative" approach (Horvitz, 1999) aiming to combine human and machine factors in complex educational systems with considerations of more holistic factors (Cuban, Kirkpatrick, & Peck, 2001). There is an urgent need for understanding the factors influencing the adoption of AI-based adaptive learning platforms in real-world practice. In order to fill in this gap, in this research paper we present,

1) The development of a reliable survey instrument to measure more holistic factors influencing the adoption of adaptive learning platforms in schools.

2) The results of a country-level implementation of the survey and compute the extent to which they predict teacher engagement with adaptive learning platforms.

Based on the results observed, we provide suggestions for scaled adoptions of AI-based adaptive learning platforms in schools more widely.

## 2      Context and the Adaptive Learning Platform Studied

Adoption of AI-based adaptive learning platforms, or any particular artefact as a matter of fact, is first and foremost likely to be influenced by the particular features of the artefact itself. Hence, building upon Vandewaetere and Clarebout's (2014) four-dimensional framework to structure the diversity of adaptive learning platforms in schools, we first present the details of the particular adaptive learning platform we investigated. The specific adaptive platform we studied is called Anonymised-for-review, which is a student-centred adaptive learning system developed by Anonymised for review. The platform has been implemented and used as the primary learning resource across six core curricula from Grade 5 to Grade 12 in all K-12 public schools in Anonymised. By design, the Platform is a student-centred adaptive learning system which allows learners to self-regulate learning through adaptive tests, bite-sized multimedia content and analytics that provides feedback on cognitive and behavioural performance. Adaptation is dynamic and covers both the content and the feedback. The Platform allows shared control from both students and teachers. So, teachers can control lessons assigned to students for classroom management, curriculum pacing and behavioural management purposes. In addition, students, who lack training in self-regulated learning and struggle to engage with adaptive content, can override the platform's suggestions. By providing control and continuous analytics feedback about what students are working on and how they are performing, teachers can support, and intervene at the right point in a student's learning process in classrooms regardless of system suggestions. Additionally, school leaders are also provided with learning analytics dashboards to monitor school-level performance, identify gaps for intervention and support teachers' weekly professional development. Currently, the Platform only adapts to cognitive and behavioural learner parameters based on student answers to items and engagement behaviours but does not consider any affective or meta-cognitive dimensions in adaptation.

## 3      Methodology

Based on previously established methodologies published (i.e Nazaretsky et al., 2022), the development process included the following five steps: 1) Top-down design of the categories and items based on the literature review of factors influencing technology adoption in schools and bottom-up design of additional items based on discussions with teachers and experts, (2) Pilot with a small group of teachers for polishing the initially designed items, (3) Exploratory Factor Analysis for bottom-up analysis of emerged factors, (4) Reliability analysis of the emerged factors to verify their internal validity.

In the first step, we have undertaken a semi-systematic literature review on factors that influence the adoption of AI in schools and similar to Debeer *et al*., (2021) we run the following search string on the web of science database (TS=((adaptiv* OR adjust* OR personal* OR individual* OR tailor* OR custom* OR intelligent* OR tutor*) AND (learn* OR education* OR class* OR school OR elementary OR primary OR secondary) AND (digital* OR online OR computer*) AND (adopt* OR use* OR interve* OR experiment*))) and we filtered papers from the last 25 years. Based on the review of the identified papers, it became clear that the change connected to the adoption of AI in schools does not merely happen by means of introducing a new digital tool for learning at the classroom level, rather, it is profoundly connected to existing teaching practices, school leadership, teachers' vision and perception of the platform, as well as the infrastructure and technical pedagogical support. In this respect, in order to understand and measure factors influencing AI adoption in schools, the need for an instrument that counts for the different dimensions of schools that are engaged when innovative transformations take place became clear. Based on the summary of the identified papers we categorised indicators of adoption at the school level in four: a) digital maturity infrastructure, b) pedagogical, c) governance and d) teacher empowerment and interaction dimensions. These dimensions are selected because they broadly cover the essential elements relevant to the adoption of AI in schools, as discussed in research studies of the relevant literature. Secondly, unlike available technology adoption instruments in the literature, these aspects do not *only focus* on the technical aspects of the platform features, and they are not only at an individual teacher level. Rather they consider multidimensional factors at the whole-school level which are crucial to the successful adoption of AI-based Adaptive Platforms in Schools. These aspects involve conceptual constructs of the framework from which the initial instrument items were built. It is important to note that these four aspects should not be perceived as isolated entities that do not transverse and correlate. Rather, these aspects are interrelated and influence one another.

To summarise the identified dimensions briefly,

a) Digital maturity is understood as an aspect that sheds light on the organizational aspect of digital adoption, technical infrastructure and teachers' knowledge and confidence in the use of adaptive learning platforms. Adopting a whole-school approach towards technology implementation necessitates examining the level at which the school as an organizational entity is already integrated with digital technologies (e.g., the infrastructure, processes and interactions). Moreover, it is crucial to evaluate the level at which the staff, including teachers and leaders trust, feel at ease and confident with using and deploying AI tools in their daily practices.

b) The pedagogical aspect mainly aims to get a sense of to what extent adaptive learning platforms are actively embedded within the pedagogical practices of the school and require changes from the traditional practices of teacher's practice. This dimension is also associated with the school's curriculum alignment, teachers' workload, and assessment cultures of the schools. The governance dimension requires an account of the school as a structural organization in which a shared vision about the role of AI tools is put into practice.

c) The governance is particularly significant, as with any organizational change needing to be stabilized and supported at the management level. In this respect, the governance aspect sheds light on the vision of different school actors on the use of adaptive learning platforms and more specifically focuses on the vision of school leaders and principals about the direction they envisage for their school with respect to the adoption of AI tools.

d) At last, the teacher empowerment and interaction dimension aims to generate insights into the extent to which individual teachers help each other out and create learning communities for sharing and interacting with their experiences of a particular AI tool. Moreover, this aspect revisits the role of students as active contributors to adoption through shared school-level practices.

After the literature review was completed, we ran in-depth discussions and interviews with the creators of the platform to cross-check survey items to be aligned with the key design principles for the platform as well as their professional experience in implementing adaptive learning platforms in schools. Then, the initial set of items was piloted with a group of 10 experienced teachers. The participants were informed about the purpose of the survey and its development methodology, then asked to fill in the survey individually and discuss any feedback they considered relevant. As a result, the survey items were finalized and translated as required. Relevant ethical and practical approvals were obtained from school principals and local education governing agencies before large-scale implementation. The resulting instrument consisted of 30 five-level Likert items to measure potential factors influencing the adoption of adaptive learning platforms in K12 public schools and 8 items on socio-demographics (The instrument can be seen in the Appendix below).

To explore the factorial structure of the collected data all items were subjected to exploratory factor analysis (EFA) using the principle factor extraction (PFE) approach. Since our theoretical and experimental results indicated correlations between potential factors we chose a dimensionality reduction method that does not assume that the factors are necessarily orthogonal (i.e principal axis factoring). The analysis was completed using relevant packages in *Python*. Initially, data was examined for outliers and erroneous responses were removed, such as respondents with incorrect email addresses or invalid ages. Then, for the remaining 529 responses, the Kaiser-Meyer-Olkin (KMO) Measure of Sampling Adequacy (MSA) was used to verify the sampling adequacy for the analysis (KMO = 0.913 > 0.9). Bartlett's test of sphericity ($\chi^2$ = 9819.2, $p < 0.001$) indicated that the correlation structure is adequate for meaningful factor analysis. Using the *FactorAnalyzer* package we ran the EFA with the PFE approach and the results showed seven factors with eigenvalues over 1. The seven-factor model also met all the quality criteria suggested by Bentler and Bonett (1980) and was chosen as the best model for explaining the response data. For deciding on the mapping of items to factors, we used a threshold of .75. We didn't have any items with double loading thanks to the high threshold set and items with no loading (not explained by any of the factors) were qualitatively judged to decide whether they should be retained or dropped. Then, we examined the internal consistency of the assignment of items to a factor and labelled each factor accordingly. The factors identified and labelled are presented below.

**Factor 1.** presented strong positive correlations with 'No Additional Analytics', 'No Additional Switching Tools', 'No Additional Classroom Behavioural Management', 'No Additional Balancing of Learning' items which are all related to not adding any additional workload to teachers. So, we labelled this factor as the workload factor.

**Factor 2.** showed strong positive correlations with 'Believe in Success', 'Trust in the Platform', 'Defending the Platform' items which are all related to measuring the extent to which teachers have ownership of the platform and trust it. We labelled this factor as teachers' ownership factor.

**Factor 3.** had strong positive correlations with 'Part Of Community', 'Helpful Professional Development', 'Access To Help', 'Sufficient Guidance' items which all relate to the extent to which teachers feel that they are supported and get help from a community. Therefore, we labelled this factor as teachers' perceived support factor.

**Factor 4.** had strong positive correlations with 'Knowledge And Skills To Use' and 'Self-Efficacy' items so we labelled this factor as teachers' perceived knowledge and confidence factor.

**Factor 5.** showed strong positive correlations with 'Maintaining Engagement', and 'Finding Balance' items which are related to teachers' sense of control in the use of the platform in their classroom practice. So, we labelled this factor as the classroom orchestration factor.

**Factor 6.** presented strong positive correlations with 'Great Content', 'Efficiency', and 'Satisfaction with the Platform' items which all relate to teachers' perceived quality of digital learning content and platform features. Therefore, we labelled this product quality factor.

**Factor 7.** had a strong negative correlation with 'No Privacy Concerns' item so we labelled it as the ethical factor.

**Table 1. Seven Factors and Variance Explained**

| Factors | Sum of Squared Loadings | Average Var Extracted | Cumulative Variance |
|---|---|---|---|
| Factor 1 | 3.270 | 0.093 | 0.093 |
| Factor 2 | 3.119 | 0.089 | 0.183 |
| Factor 3 | 2.915 | 0.083 | 0.266 |
| Factor 4 | 1.997 | 0.057 | 0.323 |
| Factor 5 | 1.894 | 0.054 | 0.377 |
| Factor 6 | 1.519 | 0.043 | 0.420 |
| Factor 7 | 1.172 | 0.033 | 0.454 |

## 4    Results

### 4.1 Teachers' responses to the Items

The instrument was double-translated and delivered bilingually through Survey Monkey. It was administered to 792 public school teachers from Grade 5 to Grade 12, teaching 6 subjects including English, Math, and Science. Teachers' age ranged from 22 to 55, with the average age of 44. Among the teachers, only 2% were novice teachers with 1-2 years of teaching experience while above 80% of teachers had at least 6 years of teaching experience. With regards to their experience using adaptive learning platforms, 6.4% had less than a year's experience, 13% had 1-2 years of experience, 30% had 2-3 years of experience and about 50% had 3-5 years of experience. The responded teachers consisted of a balanced gender distribution. In terms of results, we first used the instrument to measure the distribution of teachers' responses to each of the factors that resulted from the analysis. Per factor, the score of the teacher on that factor was computed by averaging the scores of the items belonging to that factor. The score of the individual item regarding the ethical concern was taken as it is. The resulting distributions of the scores by the chosen subcategories are presented in Figure 1.

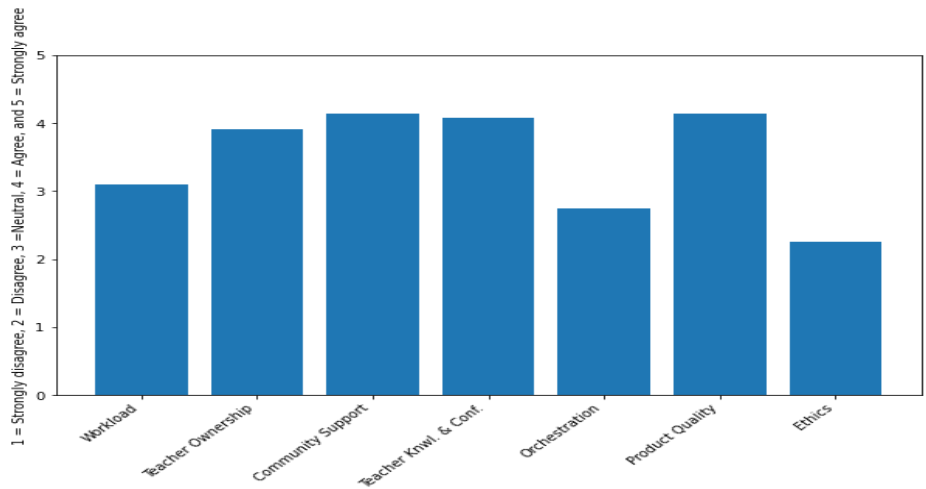

**Fig. 1.** Teachers' mean responses to seven factors of the survey.

Based on the mean values provided, the mean value for the workload factor was 3.09, indicating that the teachers didn't think that the adaptive learning platform added to or decreased their workload. The mean value for the teacher ownership factor was 3.92, 4.13 for the community support factor, and 4.07 for the teacher knowledge and confidence factor. These indicate that teachers overall agreed that they had enough amount of knowledge and confidence to use the adaptive learning platform, they receive enough support for its effective use and they have high amount of ownership of the platform. The mean value for the product quality factor was 4.14, indicating that teachers on average agreed that the platform was of good quality and were satisfied with it. On the other hand, the mean value for the orchestration factor was 2.74, indicating a somewhat disagreement that teachers can find a balance between the other activities of teaching and the use of the platform as well as struggling to keep students engage with the platform during classroom activities. Finally, the mean value for the ethics factor was 2.25, indicating that teachers might have some concerns about privacy, safety, and data security while using the platform.

### 4.2 Predicting teachers' engagement with the adaptive learning platform

Furthermore, we studied the correlations between the items included in the seven factors identified above and teachers' active days on the adaptive learning platform. Adaptive learning days were calculated using the platform's log data for the first half of the 2022-2023 academic year. As summarized in Table 2, most items that correlated statistically significantly with teachers' engagement with the adaptive learning platform were related to the seven factors discussed above. Those that are different are the items that had double loadings on multiple factors. For instance, the extent to which teachers think that the platform provides 'Helpful Feedback' is related to both ownership and community support factors, or 'content alignment' with national curricula item is related to ownership and product quality factors, 'Easy of Use' item is loaded on ownership, community and quality factors.

**Table 2. Significantly correlating items with teachers' real-world engagement with the adaptive learning platform**

| | Pearson Correlation Coefficient | P-Value | Adjusted-P-Value |
|---|---|---|---|
| **Helpful CPD** | 0.246061 | p = 0.00 < .001 | p = 0.00 < .001 |
| **Believe in Success** | 0.225836 | p = 0.00 < .001 | p = 0.00 < .001 |
| **Satisfied w/ Platform** | 0.200222 | p = 0.00 < .001 | p = 0.00 < .001 |
| Helpful Feedback | 0.190905 | p = 0.00 < .001 | p = 0.00 < .001 |
| **Defend the Platform** | 0.190418 | p = 0.00 < .001 | p = 0.00 < .001 |
| **Great Content** | 0.182100 | p = 0.00 < .001 | p = 0.00 < .001 |
| **Platform is Efficient** | 0.180516 | p = 0.00 < .001 | p = 0.00 < .001 |
| **Trust in the Platf.** | 0.178809 | p = 0.00 < .001 | p = 0.00 < .001 |
| **Part of Community** | 0.154447 | p = 0.00 < .001 | p = 0.00 < .01 |
| Content Alignment | 0.153034 | p = 0.00 < .001 | p = 0.00 < .01 |
| Easy to Use | 0.144989 | p = 0.00 < .001 | p = 0.00 < .01 |
| **Sufficient Guidance** | 0.142695 | p = 0.00 < .001 | p = 0.00 < .01 |
| Leadership's Believe | 0.139671 | p = 0.00 < .01 | p = 0.00 < .01 |
| **Access to Help** | 0.134303 | p = 0.00 < .01 | p = 0.00 < .01 |
| **Can Find Balance** | 0.131564 | p = 0.00 < .01 | p = 0.01 < .01 |
| **No Ext. Lesson Planning** | 0.112758 | p = 0.01 < .01 | p = 0.02 < .05 |
| **Self-Efficacy** | 0.108566 | p = 0.01 < .05 | p = 0.03 < .05 |
| Opportunity to Share | 0.106320 | p = 0.01 < .05 | p = 0.03 < .05 |

At last, we used the data from the survey to build a predictive model of teachers' engagement with the adaptive learning platform in their real-world practice. More specifically, we used the *catboost* algorithm with the following parameters: 1000 iterations, depth of 2, the learning rate of 0.01, and early stopping rounds set to 30. These parameters were chosen based on previous literature and our experiments to optimize the model's performance. The results indicated an $R^2$ value of 0.244, showing that the model was able to explain %24 of the variation in teachers' real-world engagement data. Although there is plenty of space to improve the model's predictive power, for complex social phenomena like teachers' real-world engagement with adaptive learning platforms in schools, the results can be considered promising. Perhaps, more importantly, the highest co-efficiency values were associated with the extent to which teachers perceived professional development provided as useful and their belief in the potential success of the platform to improve learning which further highlights the significance of these items for predicting teachers' engagement with the adaptive learning platform.

## 5    Discussion

The use of AI in the field of education has greatly contributed to the design and development of effective adaptive learning platforms. However, as these platforms move out of our labs to schools, most researchers have come to realize that many important and complex problems regarding their adoption exceed the scope of the system features. Change connected to any innovation adoption does not merely happen by means of introducing a new digital tool for learning at the classroom level, rather, it is profoundly connected to teaching practices, school leadership, teachers' vision and perception as well as the infrastructure and technical pedagogical support (Ilomäki & Lakkala, 2018). Recent research in AIED literature highlights the importance of such factors and shows that the combination of human mentoring and AI-driven computer-based tutoring can have a positive impact on student performance, with several studies demonstrating the promise of this approach (Chine *et al.*, 2022). However, teachers' acceptance level of AI technologies influences the integration of such human-AI teaming approaches (Ifinedo et al., 2020). In this respect, for human-AI personalisation approaches to work, there is a need for an approach towards innovation adoption that accounts for the different dimensions of schools that are relevant when innovative transformations take place. To scale the adoption of adaptive learning platforms, a multitude of other relevant factors should work together to encourage successful implementations in schools. In this study, we developed a reliable instrument to measure some of these important factors in the adoption of adaptive learning platforms and built a predictive model of teachers' real-world engagement using the survey data as input.

Our results highlighted a set of influential factors that play an important role in the adoption of adaptive learning solutions by teachers. First of all, it is important that the adaptive platform does not lead to an increase in teachers' existing workload (even if it does not lead to any decrease in it). Teachers' workload is already a significant concern in many education systems and any adaptive platform system implementation should aim to protect status quo or decrease the workload rather than adding to it by any means. Our results indicate that no additional workload to use learning analytics, no additional workload of switching between different tools, no additional workload of classroom management, and no additional workload of switching between different pedagogical practices during the implementation of the adaptive platforms is desired. Therefore, understanding teachers' existing practices in schools and thinking about when and how the use of adaptive platforms should be implemented requires careful thinking. Second, a significant amount of effort should be put into increasing teachers' trust and ownership of the adaptive platform. The more teachers believe in the potential of the platform and trust in its success, the more likely they are to adopt it in their practice. Involving teachers in the iterative research and design process of the adaptive learning platform, and making sure it is genuinely addressing teacher needs might help with teacher trust and ownership (Holstein *et al.,* 2019). In addition, the extent to which individual teachers receive guidance, professional development, and support regarding the use of the adaptive platform as well as the increased opportunities for teachers to help each other and create learning communities for sharing and interacting with their experiences of the platform appears to play a significant role in their likelihood of engaging with the adaptive learning platforms. As well-evidenced in previous work regarding the adoption of technology in education in general, ease of use, the perceived quality of the platform as well as teachers' skills and confidence in using them were also identified as key elements in gaining and maintaining teachers' engagement with adaptive learning platforms. In addition, the dimensions of the orchestration of the platform in classroom settings and privacy concerns were highlighted in our results. Therefore, designing lesson plans that help teachers to more effectively switch between the use of the adaptive platform and other pedagogical activities, providing teachers guidance on how to increase their students' engagement with the platform during class time as well as ensuring that the platform does not cause any ethical and privacy-related concerns can increase the teachers' adoption of adaptive learning platforms.

It was interesting to observe that the importance and need for strong investment in the technical infrastructure to ensure that teachers have access to reliable hardware, software, and support when they are implementing the adaptive learning platforms were not specifically highlighted in our results. The significance of such factors associated with the infrastructure is well-reported in the literature (Ifinedo et al., 2020). However, this surprising result is likely due to the fact that in the context of our study, for schools we studied there is a system-level top-down approach to digitising teaching and learning, providing schools with all the material aspects of technology needed (e.g. laptops, charging, storage, internet, etc.) as well as the capacity within schools to communicate potential technical challenges are part of the product and services offered.

At last, our prediction model of teachers' real-world engagement with the adaptive learning platform achieved an $R^2$ value of 0.244. This is a humble but promising result to predict a complex and dynamic phenomenon like teachers' real-world engagement. AI has proven its potential to provide applications for adaptive learning to support students' academic performance (VanLehn, *et al.,* 2020), yet another valuable potential of AI models is to help us increase the dimensions of variability in our system-level models of complex constructs like the adoption of adaptive platforms. The results of this study not only can contribute to practices of wider adoption of adaptive learning platforms in schools through the suggestions provided above, but also can help us build highly dimensional models of predicting teachers' real-world engagement behaviours.

How to promote the broad adoption of AI in Education is one of the main goals of our community and an important step we need to take to progress in our mission of using AI to contribute to a world with equitable and universal access to quality education at all levels (UN's SDG4). If we are to scale the use and adoption of adaptive learning platforms in schools, we need to better understand, measure and model system-level variables identified in this study.

## 6 Conclusion

The scaled adoption of adaptive learning platforms has the potential to revolutionize the way students learn and teachers teach. While previous research has predominantly focused on developing technologies that are both pedagogically and technologically sound, the factors influencing real-world adoption have received limited attention. To the best of our knowledge, there is no previously established instrument for measuring school-level factors influencing the adoption of adaptive platforms. Hence, the contribution of this research paper is twofold. First, it introduces a new instrument to measure school-level factors influencing teachers' adoption of adaptive platforms in schools and presents seven factors with their internal reliability and validity. Second, it presents a large-scale implementation of the survey and develops a predictive model of teachers' real-world engagement using this survey data. These improve our understanding of the factors influencing the adoption of adaptive learning platforms in schools and can be used to increase further adoption of AI in real-world school settings.

## Appendix

The full survey used can be accessed here.